\documentstyle[preprint,aps,eqsecnum]{revtex}

\tighten
\begin{document}
\draft
\preprint{JIHIR 93-10}

\title   {
                Approximate Particle Number Projection for Rotating Nuclei
         }
\author  {                  P. Magierski, S. \'Cwiok
         }
\address {   Institute of Physics, Warsaw University of Technology,\\
                  ul. Koszykowa 75, PL--00-662 Warsaw, Poland
         }
\author  {                        J. Dobaczewski
         }
\address {    Institute of Theoretical Physics, Warsaw University,\\
                     ul. Ho\.za 69, PL--00-681 Warsaw, Poland
         }
\author  {                        W. Nazarewicz\thanks{
                Permanent address: Institute of Theoretical Physics,
                Warsaw University, Ho\.za 69, 00-681 Warsaw, Poland;
                Institute of Physics, Warsaw University of Technology,
                Warsaw, Poland.}
         }
\address {           Joint Institute for Heavy-Ion Research,
                        Oak Ridge National Laboratory, \\
                   P.O. Box 2008, Oak Ridge, TN37831, U.S.A.\\
                                      and\\
                 Department of Physics, University of Tennessee,\\
                           Knoxville, TN 37996, U.S.A.
         }

\maketitle

\begin{abstract}
Pairing correlations in rotating nuclei are discussed within the
Lipkin-Nogami method. The accuracy of the method is tested for
the Krumlinde-Szyma\'nski R(5) model. The results of
calculations are compared with those obtained from the standard
mean field theory and particle-number projection method, and
with exact solutions.
\end{abstract}

\narrowtext
\section{Introduction}

The mean-field approach to the phenomenon of superconductivity
introduced by Bardeen, Cooper, and Schrieffer \cite{Bar57}
({\sc bcs}) allows for a simple and elegant treatment of pairing
correlations in nuclei \cite{Boh58,Bel59}.  The main drawback of
the {\sc bcs} method is that its wave function is not an
eigenstate of the particle number operator.  The accuracy of the
{\sc bcs}  approximation is satisfactory if the pairing
interaction strength is strong or the number of particles is
very large \cite{Lan64,RS}.  These conditions are  not satisfied
in nuclei.  Indeed, the critical value of the effective pairing
strength, $G_{\text{crit}}$, above which the static gap exists,
is inversely proportional to the single-particle level density
around the Fermi level \cite{Bel59}, i.e., it becomes very large
around subshell closures.

It was demonstrated  by Lipkin \cite{Li} that the effect of the
nucleon number fluctuation can be suppressed by using a model
Hamiltonian $ \hat{H}-\lambda_{1}\hat{N}-\lambda_{2}\hat{N}^{2}
$ instead of the conventional Routhian $\hat{H}-\lambda \hat{N}
$, where $\hat{H} $ is the original  Hamiltonian (involving
pairing interaction) and $\hat{N} $ is the nucleon number
operator. The approach by Lipkin was then developed by Nogami
and his collaborators \cite{No64,NZ,No65,GN,PNL}.  The important
feature of the Lipkin-Nogami ({\sc ln})  method is that (i)
there always exists a nontrivial (superfluid) solution
regardless of the strength of pairing force and (ii) the
{\sc ln} wave function has a similar form to that of the
{\sc bcs} method, thus allowing a simple interpretation of
excited states in terms of quasiparticles.

In this respect it seems to be of a considerable interest to
extend the {\sc ln} method to the case of rotation, where the
short-ranged attractive pairing force plays a significant role.
For instance, at low spins the pairing  correlations tend to
significantly reduce the nuclear moment of inertia as compared
to the rigid-body estimate. On the other hand, it is well known
that at very high spins many nuclei behave as macroscopic
rotors, i.e. their moments of inertia are fairly constant and
close to their rigid body values.  More examples illustrating
the importance of pairing correlations at high spins can be
found in
the review in Ref. \cite{Shi89}.

In this connection further studies of the Mottelson-Valatin
effect \cite{MV} (phase transition from superfluid to normal
state in rotating nuclei) are of great importance
\cite{NDS,Shi89}. Since the {\sc bcs}  method provides a rather
poor description of the pairing phase transition region, one
expects that the {\sc ln} method would be a powerful tool
allowing for a better description of pairing correlations
without losing the simplicity of the rotating independent
quasiparticle picture.

In Sec. 2  the cranked {\sc ln} ({\sc lnc}) equations are
derived.  The equation for the particle number fluctuation is
explicitly written in terms of the single-particle and pair
densities. This will be useful when adopting the {\sc ln}
approach to the general Hartree-Fock-Bogolyubov ({\sc hfb})
treatment.
The method is  applied in Sec. 3 to the exactly solvable
two-level cranking model with pairing.  The exact solutions are
compared to those obtained within the standard mean field
approach, particle number projection method, {\sc lnc}
treatment, and its particle-number projected version.  The
conclusions are presented in Sec. 4.

\section{The method }
\subsection{Cranked Lipkin-Nogami  equations }

Let us consider the {\sc bcs} Hamiltonian which contains a
single-particle Hamiltonian, $\hat H_{\text{sp}}$, and a
seniority pairing (monopole, state independent)  interaction:
\begin{equation}
\hat{H}=\hat{H}_{\text{sp}}+\hat{H}_{\text{pair}} =
\sum_{k}e_{k}a^{+}_{k}a_{k}
        - G\sum_{k,l>0}a^{+}_{k}a^{+}_{\bar{k}}a_{\bar{l}}a_{l},
\end{equation}
where $e_{k}$ is the single-particle energy, $G$ is the pairing
strength, and $|\bar{k}\rangle =\hat T|k\rangle $.

In order to investigate the pairing interaction in rotating
system we consider the cranking Hamiltonian (Rout\-hian):
\begin{equation}\label{routhian}
\hat{H}^{\omega}=\hat{H}_{\text{sp}}
                +\hat{H}_{\text{pair}}-\omega\hat{j}_{x},
\end{equation}
where $\hat{j}_{x}$ denotes the component of the total nucleonic
angular momentum on the rotational axis (here: x--axis). This
axis is assumed to be fixed in space (only one-dimensional
rotation is considered) and the angular velocity of rotation,
$\omega$, is supposed to be constant.

In the standard rotating {\sc bcs} ({\sc rbcs}) approach the
expectation value of the Routhian (\ref{routhian}),
$E^{\omega}=\langle\hat{H}^{\omega}\rangle$,  is minimized in
the product state of independent quasiparticles defined through
the Bogolyubov transformation:
\begin{eqnarray}
\left\{
\begin{array}{crc}
\alpha_{i}^{+} &=&\displaystyle{\sum_{l}}
                  (A_{li}a^{+}_{l}+B_{li}a_{l}),\\
\alpha_{i}     &=&\displaystyle{\sum_{l}}
                  (A_{li}^{*}a_{l}+B_{li}^{*}a_{l}^{+}),
\end{array}
\right.
\end{eqnarray}
where $\alpha^{+}_{i} (\alpha_{i} )$ is the quasiparticle
creation (annihilation) operator.  The {\sc rbcs} wave function
represents a mixture of states with different numbers of
particles. Consequently, in order to account for the
fluctuations, the particle number should be projected before
variation.  This can be done in a good approximation by means of
the {\sc ln} method outlined below.

Let us assume that the state $|\psi_{n}\rangle $ is the
quasiparticle vacuum, i.e.,
\begin{equation}
\alpha_{i}|\psi_{n}\rangle =0.
\end{equation}
The index $n$ stands for the average number of particles in the
state $|\psi_{n}\rangle $, determined by means of the particle
number equation
\begin{equation}\label{pne}
\langle\psi_{n}|\hat{N}|\psi_{n}\rangle =n,
\end{equation}
where $\hat{N}=\displaystyle{\sum_{k}a^{+}_{k}a_{k}}$ is the
particle number operator. The state $|\psi_{n}\rangle $ can be
expanded in eigenstates of the particle number operator,
\begin{equation}
|\psi_{n}\rangle = \sum_{n_{0}}c_{n,n_{0}}|\phi_{n_{0}}\rangle,
\end{equation}
\begin{equation}
\hat{N}|\phi_{n_{0}}\rangle = n_{0}|\phi_{n_{0}}\rangle.
\end{equation}
The total Routhian (\ref{routhian}) commutes with the particle
number operator, $[\hat{H}^{\omega},\hat{N}]=0$.  By expanding
$E^{\omega}$ in terms of $n$
\begin{equation}\label{rouex}
\langle\phi_{n}|\hat{H}^{\omega}|\phi_{n}\rangle = \lambda (n)=
\lambda_{0}+\lambda_{1}n+\lambda_{2}n^{2}+...
\end{equation}
the following relation is obtained:
\FL
\begin{eqnarray}  \label{expan}
\langle\phi_{n_{0}}|\hat{H}^{\omega}|\phi_{n_{0}}\rangle &=&
\langle\psi_{n}|\hat{H}^{\omega}|\psi_{n}\rangle -
\lambda_{1}(\langle\psi_{n}|\hat{N}|\psi_{n}\rangle - n_{0})
                                                \nonumber \\
&-&\lambda_{2}(\langle\psi_{n}|\hat{N}^{2}|\psi_{n}\rangle -
n_{0}^{2})-...
\end{eqnarray}
The important feature of the above expression is that it yields
the expectation value of $\hat{H}^{\omega}$ in the projected
{\sc rbcs} state, $|\phi_{n_{0}}\rangle$, in terms of the
expectation value of Routhian in the {\sc rbcs} state,
$|\psi_{n}\rangle$. This implies that, knowing coefficients
$\lambda_{i}$, one can minimize the right hand side of
Eq.{{\ }}(\ref{expan}) instead of minimizing explicitly the
expectation value of $\hat{H}^{\omega}$ in the projected
{\sc rbcs} state.

So far the considerations are exact.  In the next step, however,
the expansion (\ref{expan}) is truncated by retaining the first
$m$ terms; the coefficients
$\lambda_{1},\lambda_{2},...,\lambda_{m}$ are then calculated
from the following set of linear equations:
\begin{eqnarray} \label{lameq}
\left\{
\begin{array}{crc}
\langle\psi_{n}|\hat{K}^{\omega}\hat{N}|\psi_{n}\rangle &=&
\langle\psi_{n}|\hat{K}^{\omega}|\psi_{n}\rangle\langle
\psi_{n}|\hat{N}|\psi_{n}\rangle \\
\langle\psi_{n}|\hat{K}^{\omega}\hat{N}^{2}|\psi_{n}\rangle &=&
\langle\psi_{n}|\hat{K}^{\omega}|\psi_{n}\rangle\langle
\psi_{n}|\hat{N}^{2}|\psi_{n}\rangle \\
......................&...&............................. \\
\langle\psi_{n}|\hat{K}^{\omega}\hat{N}^{m}|\psi_{n}\rangle &=&
\langle\psi_{n}|\hat{K}^{\omega}|\psi_{n}\rangle\langle
\psi_{n}|\hat{N}^{m}|\psi_{n}\rangle ,
\end{array}
\right.
\end{eqnarray}
where
$\hat{K}^{\omega}=\hat{H}^{\omega}-\lambda_{1}\hat{N}-\lambda_{2}
\hat{N}^{2}-...-\lambda_{m}\hat{N}^{m}$. The  $m=1$ case, together
with the constraint (\ref{pne}), is equivalent  to the standard
{\sc rbcs} method.  The case of $m$=2 is discussed below. The
generalization to higher orders is straightforward although
algebraic manipulations become tedious. Since the right hand
side of equation (\ref{rouex}) has to be finite, the $i$-th
order correction to Routhian, which is proportional to
$\lambda_{i}$ is of the order of $n^{-i}$ .  Therefore, when one
deals with a system with large number of particles it seems
reasonable to neglect the third order term (the mass formula
derived from the seniority scheme is actually quadratic in $n$).

In the {\sc ln}  method the expectation value of
$\hat{K}^{\omega}$ is minimized assuming that coefficients
$\lambda_{1}$ and $\lambda_{2}$ are constant (the simultaneous
variation of $\lambda$'s would lead to more complicated
equations \cite{RS}):
\begin{equation}\label{var}
\delta\langle\psi_{n}|\hat{K}^{\omega}|\psi_{n}\rangle = 0.
\end{equation}
The Lagrange multipliers $\lambda$ and $\omega$ are determined
by fixing expectation values of the particle number and angular
momentum, respectively,
\begin{equation}
\langle\psi_{n}|\hat{N}|\psi_{n}\rangle = n,
\end{equation}
\begin{equation}
\langle\psi_{n}|\hat{j}_{x}|\psi_{n}\rangle = I.
\end{equation}
It is convenient to express the results through the density
matrices $\rho$ and $u$ and the  pairing tensor $\chi$, which
are defined by means of the transformation matrices $A$ and $B$:
\begin{equation}\label{rhoandu}\begin{array}{rcl}
\rho_{kl}&=&\sum_{i}B_{ki}^{*}B_{li}, \\
u_{kl}&=&\sum_{i}A_{ki}^{*}A_{li}=\delta_{kl}-\rho_{kl}, \\
\chi_{kl}&=&\sum_{i}A_{li}B_{ki}^{*}.
\end{array}\end{equation}
Expressed in terms of quasiparticles the operator
$\hat{K}^{\omega}$ takes the form
\begin{equation}
\hat{K}^{\omega}={K}^{\omega}_{00}+\hat{K}^{\omega}_{11}+
                   \hat{K}^{\omega}_{20}+\hat{K}^{\omega}_{22}+
\hat{K}^{\omega}_{31}+\hat{K}^{\omega}_{40},
\end{equation}
where
\FL
\begin{eqnarray}
{K}^{\omega}_{00}&=&\sum_{k,l}\left[e_{k}\delta_{kl}-
 \omega (j_{x})_{kl}\right]\rho_{kl}-\frac{\Delta^{2}}{G}-
\frac{1}{2}G\sum_{k,l}\rho_{kl}\rho_{\bar{k}\bar{l}}
                                                \nonumber \\
&-&\lambda_{1}n-\lambda_{2}n^{2}-2\lambda_{2}\sum_{k,l}
 \rho_{kl}u_{kl},
\end{eqnarray}
\FL
\begin{eqnarray}
\hat{K}^{\omega}_{11}&=&\sum_{i,j}\Bigg{\{}\Bigg{[} \sum_{k,l}
(\epsilon_{kl}- \lambda-\omega
(j_{x})_{kl})(A_{ki}^{*}A_{lj}-B_{kj}B_{li}^{*}) \nonumber \\
&+&\Delta_{kl}A_{ki}^{*}B_{lj}- \Delta_{kl}^{*}B_{ki}^{*}A_{lj}
\Bigg{]}+\lambda_{2}\delta_{ij}\Bigg{\}}\alpha_{i}^{+}\alpha_{j},
\end{eqnarray}
\FL
\begin{eqnarray}
\hat{K}^{\omega}_{20}&=&\sum_{i,j}\Bigg{\{}
 \sum_{k,l}\left[\epsilon_{kl}-
 \lambda-\omega (j_{x})_{kl}\right]A_{ki}^{*}B_{lj}^{*}
                                                  \nonumber \\
&+&\frac{1}{2}\Delta_{kl}A_{ki}^{*}B_{lj}^{*}+
\frac{1}{2}\Delta_{kl}^{*}
B_{li}^{*}B_{kj}^{*}\Bigg{\}}\alpha_{i}^{+}\alpha_{j}^{+}+h.c.
\end{eqnarray}
The terms $\hat{K}^{\omega}_{22}$, $\hat{K}^{\omega}_{31}$,
$\hat{K}^{\omega}_{40}$ represent the residual interaction
between quasiparticles and are neglected in this approximation.
In the above relations the following quantities are introduced:
\begin{equation}
\epsilon_{kl}=e_{k}\delta_{kl}-G\mbox{sign}(k)\mbox{sign}(l)
\rho_{\bar{k}\bar{l}}+
 4\lambda_{2}\rho_{kl}^{*},
\end{equation}
\begin{equation}
\lambda=\lambda_{1}+2\lambda_{2}(n+1),
\end{equation}
\begin{equation}
\Delta_{kl}= -\delta_{k\bar{l}}\mbox{sign}(k)\Delta=
  -G\delta_{k\bar{l}}\mbox{sign}(k)\sum_{k>0}\chi_{k\bar{k}}.
\end{equation}
Condition (\ref{var})  leads to the {\sc hfb} equations:
\begin{eqnarray} \label{hfb}
\begin{array}{crc}
\displaystyle{\sum_{l}}\{(\nu_{kl}^{\omega}-
\lambda\delta_{kl})A_{li}+
 \Delta_{kl}B_{li}\} &=& {\cal E}^{\omega}_{i}A_{ki}\\
\displaystyle{\sum_{l}}\{(\nu_{kl}^{\omega*}-
\lambda\delta_{kl})B_{li}+
 \Delta_{kl}^{*}A_{li}\} &=& -{\cal E}^{\omega}_{i}B_{ki},
\end{array}
\end{eqnarray}
where
\begin{equation}
\nu_{kl}^{\omega}=\epsilon_{kl}-\omega (j_{x})_{kl},
\end{equation}
\begin{equation}\label{qrouth}
{\cal E}_{i}^{\omega}= E_{i}^{\omega}-\lambda_{2}.
\end{equation}
The difference between the usual {\sc hfb} equations and the
above ones consists in the appearance of the parameter
$\lambda_{2}$ which should be determined selfconsistently from
Eq.{{\ }}(\ref{lameq}).  It is important to note, that the
eigenvalues of the {\sc ln+hfb} equations (\ref{hfb}), ${\cal
E}^{\omega}_{i}$, are related to quasiparticle Routhians
${E}^{\omega}_{i}$ through relation (\ref{qrouth}).
Consequently, special care should be taken when interpreting the
results using the standard technique of quasiparticle diagrams
of the cranked shell model.

The total {\sc lnc}  energy of the quasiparticle vacuum is given
by
\begin{eqnarray}
E_{\text{LNC}} &=& {K}_{00}^{\omega}+\lambda_{1}n
 +\lambda_{2}n^{2}+\omega I \nonumber \\
 &=& \sum_{k}e_{k}\rho_{kk}-\frac{\Delta^{2}}{G}-
 \frac{1}{2}G\sum_{k,l}\rho_{kl}\rho_{\bar{k}\bar{l}}
                             \nonumber \\
 &-&2\lambda_{2}\sum_{k,l}\rho_{kl}u_{kl},
\end{eqnarray}
where the term proportional to $\lambda_{2}$ represents the
nucleon number fluctuation correction.

The presence of selfconsistent symmetries very often facilitates
the calculations.
One such symmetry, important in the
context of cranking model and high spins, is the signature
symmetry, i.e., the symmetry with respect to the rotation of the
system by $180^{o}$ around the x--axis. The single-particle
states with good signature $r$ are related to the original
fermionic basis by the so-called Goodman transformation
\cite{Goo74}
\begin{eqnarray}\label{Goodman}
\begin{array}{crr}
|K, r=-i\rangle &=&\frac{1}{\sqrt{2}}\left[-|k\rangle+
(-1)^{m_{k}-\frac{1}{2}}|\bar{k}\rangle\right]\\
|\tilde{K}, r=+i\rangle
&=&\frac{1}{\sqrt{2}}\left[(-1)^{m_{k}-\frac{1}{2}}|k\rangle+
|\bar{k}\rangle\right],
\end{array}
\end{eqnarray}
where $m_{k}$ is the projection of the single-particle angular
momentum on the symmetry axis.  Application of transformation
(\ref{Goodman}) leads to immediate selection rules for  the
coefficients of the Bogolyubov transformation,
\begin{equation}
A_{K\tilde{L}}=A_{\tilde{K}L}=B_{KL}=B_{\tilde{K}\tilde{L}}=0,
\end{equation}
and, consequently, for the matrix elements of the {\sc ln+hfb}
Routhian \cite{Goo74}.  In the following, it will be assumed
that the system is invariant with respect to the signature
symmetry.

\subsection{Calculation of $\lambda_{2}$ }

In order to compute $\lambda_{2}$ one can use the set of
equations (\ref{lameq}) which in the case of {\sc lnc} is
reduced to the two relations:
\begin{eqnarray}
\left\{
\begin{array}{crc}
\langle\psi_{n}|\hat{K}^{\omega}\hat{N}|\psi_{n}\rangle &=&
\langle\psi_{n}|\hat{K}^{\omega}|\psi_{n}\rangle\langle
\psi_{n}|\hat{N}|\psi_{n}\rangle \\
\langle\psi_{n}|\hat{K}^{\omega}\hat{N}^{2}|\psi_{n}\rangle &=&
\langle\psi_{n}|\hat{K}^{\omega}|\psi_{n}\rangle\langle
\psi_{n}|\hat{N}^{2}|\psi_{n}\rangle.
\end{array}
\right.
\end{eqnarray}
Because of the requirement $\hat{K}^{\omega}_{20}=0$, the first
equation is satisfied automatically and the second one is
simplified to
\begin{equation}\label{lameq1}
\langle\tilde{0}|\hat{K}^{\omega}|\tilde{4}\rangle
 \langle \tilde{4}|\hat{N}^{2}|\tilde{0}\rangle=0,
\end{equation}
where $|\tilde{0}\rangle$ denotes the quasiparticle vacuum and
$|\tilde{4}\rangle\langle\tilde{4}|$ is the projection operator
on the four-quasiparticle space. The matrix elements of
$\hat{K}^{\omega}$ and $\hat{N}^{2}$ that appear in
(\ref{lameq1}) are given by
\FL
\begin{eqnarray} \label{k40}
\hat{K}^{\omega}_{40}&=&
\displaystyle{\sum_{K,L,P}}
 \Bigg{\{}\frac{G}{4}B_{KP_{1}}B_{\tilde{K}P_{2}}
A_{\tilde{L}P_{3}}A_{LP_{4}} \nonumber \\
&+&\lambda_{2}B_{KP_{1}}A_{KP_{2}}B_{LP_{3}}A_{LP_{4}}\Bigg{\}}
\alpha_{P_{1}}\alpha_{P_{2}}\alpha_{P_{3}}\alpha_{P_{4}}
                             \nonumber \\
&+& {\text{h.c.}},
\end{eqnarray}
\FL
\begin{eqnarray} \label{n40}
\hat{N}^{2}_{40}&=&
\displaystyle{\sum_{M,N,Q}}
A_{NQ_{1}}^{*}B_{NQ_{2}}^{*}A_{MQ_{3}}^{*}B_{MQ_{4}}^{*}
\alpha_{Q_{1}}^{+}\alpha_{Q_{2}}^{+}
\alpha_{Q_{3}}^{+}\alpha_{Q_{4}}^{+} \nonumber \\
&+& {\text{h.c.}},
\end{eqnarray}
where we have already applied the Goodman transformation to
states with well defined signature.  By means of relations
(\ref{k40}), (\ref{n40}), and (\ref{lameq1}) one can now obtain
the expression for $\lambda_{2}$:
\widetext
\begin{equation}\label{lam2}
\lambda_{2}=\frac{G}{4}\frac{\displaystyle{\sum_{K,L>0}}
\{\chi_{K\tilde{L}}^{*}
 (\rho_{\tilde{K}\tilde{L}}^{*}+\rho_{KL})\}
\displaystyle{\sum_{K,L>0}}
\{\chi_{K\tilde{L}}(u_{\tilde{K}\tilde{L}}+u_{KL}^{*})\}-
2\sum_{K,L}(\chi\chi^{+})_{LK}
(\chi\chi^{+})_{\tilde{L}\tilde{K}}}
{\left[\mbox{Tr}
(\chi\chi^{+})\right]^{2}-2Tr(\chi\chi^{+}\chi\chi^{+})}.
\end{equation}
\narrowtext
It is easy to show that for $\omega =0$ the above relation
reduces to the well known result (see Ref. \cite{PNL}):
\begin{equation}
\lambda_{2}=\frac{G}{4}\frac{\displaystyle{\sum_{k>0}}
(u_{k}v_{k}^{3})\displaystyle{\sum_{k>0}}(u_{k}^{3}v_{k})-
\displaystyle{\sum_{k>0}}(u_{k}v_{k})^{4}}
{(\displaystyle{\sum_{k>0}}u_{k}^{2}v_{k}^{2})^2-
\displaystyle{\sum_{k>0}}(u_{k}v_{k})^{4}}.
\end{equation}
Relation (\ref{lam2}) together with equations (\ref{hfb})
completes the derivation of the {\sc lnc} equations.

\section{Results for the R(5) model}

In order to examine the accuracy of the {\sc lnc} approximation
we consider the two-level Krumlinde-Szyma\'nski R(5) model (see
\cite{KS71,KS73,Chu75,NDS,BBD}).  The Hilbert space of this
model consists of $\Omega$ $j$=$\frac{3}{2}$ multiplets.  As
shown in Fig.{{\ }}\ref{Fig1}, the single-particle levels are
split by the deformation of the average nuclear potential.  The
single-particle splitting is $2e$, i.e.,  the energy of the
upper levels (labelled as $|1\rangle$=$|m=3/2\rangle$ and $|\bar
1\rangle$=$T|1\rangle$) is $+e$ and that of the lower levels
(labelled as $|2\rangle$=$|m=1/2\rangle$ and $|\bar
2\rangle$=$T|2\rangle$) is $-e$.  We begin our analysis by
considering the half-filled (symmetric) system, i.e., the number
of particles is equal to $n$=$2\Omega$.  In the particular
version of the two-level model considered in this paper, the
Coriolis coupling between the lower levels is neglected, i.e.,
$(j_{x})_{2\bar{2}}=0$. The two-body pairing interaction is
assumed to be of the monopole type.  This simple model contains
the essential physical features of the nuclear structure
relevant to the interplay between pairing and rotational motion.
Its main advantage is that it can be solved exactly using the
Lie algebra associated with the symmetry group $R(5)$
\cite{KS73}.

In the standard notation of representing {\sc hfb} equations in
doubled dimensions \cite{RS}, the single-particle field $\nu$
has the form of a $4\times4$ matrix:
\begin{eqnarray}
\hat\nu^{\omega}=\left [
\begin{array}{cc}
\nu^{\omega}_{u} & 0 \\
0 & \nu^{\omega}_{l}
\end{array}
\right ],
\end{eqnarray}
where
\FL
\begin{eqnarray}
\nu^{\omega}_{u}=\left [
\begin{array}{cc}
-e-G\rho_{\bar{1}\bar{1}}+4\lambda_{2}\rho_{11}^{*}, &
-\omega-G\rho_{\bar{1}\bar{2}}+4\lambda_{2}\rho_{12}^{*} \\
-\omega-G\rho_{\bar{2}\bar{1}}+4\lambda_{2}\rho_{21}^{*}, &
e-G\rho_{\bar{2}\bar{2}}+4\lambda_{2}\rho_{22}^{*}
\end{array}
\right ],
\end{eqnarray}
\FL
\begin{eqnarray}
\nu^{\omega}_{l}=\left [
\begin{array}{cc}
-e-G\rho_{11}+4\lambda_{2}\rho_{\bar{1}\bar{1}}^{*}, &
\omega-G\rho_{12}+4\lambda_{2}\rho_{\bar{1}\bar{2}}^{*} \\
\omega-G\rho_{21}+4\lambda_{2}\rho_{\bar{2}\bar{1}}^{*}, &
e-G\rho_{22}+4\lambda_{2}\rho_{\bar{2}\bar{2}}^{*}
\end{array}
\right ].
\end{eqnarray}
In the above relation it was assumed that \cite{NDS,BBD}
\begin{equation}
(j_{x})_{21}=(j_{x})_{12}=-(j_{x})_{\bar{2}\bar{1}}=
-(j_{x})_{\bar{1}\bar{2}}=1.
\end{equation}
Similarly, the pairing field is given by
\begin{eqnarray}
\hat\Delta=-G\Omega(\chi_{1\bar{1}}+\chi_{2\bar{2}})\left [
\begin{array}{rrrr}
0&0&1&0\\ 0&0&0&1\\ -1&0&0&0\\ 0&-1&0&0
\end{array}
\right ].
\end{eqnarray}
The {\sc lnc} equations are solved iteratively with respect to
the density matrix and pairing tensor.  The initial values can
be taken from the standard {\sc rbcs} equations.  Then the
coefficient $\lambda_{2}$ is found and the matrices $A$ and $B$
are computed from Eq.{{\ }}(\ref{hfb}).  This gives a  new
approximation for the density matrix and pairing tensor.  One
should choose properly $\lambda$ (the chemical potential) at
every iteration step to satisfy the relation $n$=Tr$(\rho)$.
The above procedure is continued until the convergence is
achieved.  For more complicated Hamiltonians it is suggested to
use the so--called gradient method (see Refs. \cite{MSR,ER1,ER2}).

Having found the Bogolyubov transformation matrices $A$ and $B$,
one can calculate $\rho$ and $u$, Eq.{\ }(\ref{rhoandu}), and
then the total energy,
\FL
\begin{eqnarray}
E_{\text{tot}}&=&e \Omega (\rho_{11}+\rho_{\bar{1}\bar{1}}
-\rho_{22}-
\rho_{\bar{2}\bar{2}}) \nonumber \\ &-&\frac{\Delta^{2}}{G} -
\frac{1}{2} G \Omega \sum_{k,l>0}(\rho_{kl}
\rho_{\bar{k}\bar{l}})- 2 \lambda_{2} \Omega
\sum_{k,l>0}(\rho_{kl} u_{kl}), \label{etot}
\end{eqnarray}
the pairing energy,
\FL
\begin{eqnarray}
E_{\text{pair}}&=&E_{\text{tot}}-E_{\text{unpair}}=E_{\text{tot}}
                                                     \nonumber \\
&+&2\Omega e\sqrt{1-\Bigg{(}\frac{I}{2\Omega}\Bigg{)}^{2}}+
G\Omega\left[1-\left(\frac{I}{2\Omega}\right)^{2}\right],
\label{pair}
\end{eqnarray}
the pairing potential (energy gap),
\begin{equation}
\Delta = G\Omega (\chi_{1\bar{1}}+\chi_{2\bar{2}}),
\end{equation}
the total angular momentum,
\begin{equation}\label{itot}
I=\mbox{Tr}(j_{x}\rho)=2\Omega (\rho_{12}-\rho_{\bar{1}\bar{2}}).
\end{equation}

In the light of the recent results \cite{DN}, it is advantageous
to carry out the particle number projection after Lipkin-Nogami
({\sc lnc}+{\sc pnp}). This can easily be done by following the
formalism of Ref.{{\ }}\cite{NDS}.  The particle-number
projected energy and angular momentum are given by
\begin{equation}\label{etotpnp}
E_{\text{tot}}^{N}=E_{\text{sp}}^{N}+E_{\text{p1}}^{N}
                                    +E_{\text{p2}}^{N},
\end{equation}
\begin{equation}\label{itotpnp}
I_{x}^{N}=4\Omega\frac{\rho_{12}}{\rho_{(+)}-\rho_{(-)}}
\frac{P_{\Omega -1}(\xi)}{P_{\Omega}(\xi)},
\end{equation}
where
\begin{equation}
E_{\text{sp}}^{N}=-2\Omega e\frac{k}{\rho_{(+)}-\rho_{(-)}}
\frac{P_{\Omega -1}(\xi)}{P_{\Omega}(\xi)}
\end{equation}
is the single-particle energy, and
\widetext
\FL
\begin{equation}\label{epair1}
E_{\text{p1}}^{N}=-2G\Omega^{2}\frac{k^{2}\rho_{(+)}\rho_{(-)}}
{[\rho_{(+)}-\rho_{(-)}]^{4}}\left\{
\frac{2\rho_{(+)}\rho_{(-)}}{\Omega -1}\frac{P_{\Omega -1}'(\xi)}
{P_{\Omega}(\xi)}+[\rho_{(+)}-\rho_{(-)}]
\frac{P_{\Omega -1}(\xi)}{P_{\Omega}(\xi)}\right\}
\end{equation}
and
\FL
\begin{equation}\label{epair2}
E_{\text{p2}}^N=-G\Omega\frac{1}{[\rho_{(+)}-\rho_{(-)}]^2}
\left\{ k^2\left[ 1 -  \frac{4\rho_{(+)}^2\rho_{(-)}^2}
{(\Omega -1)[\rho_{(+)}-\rho_{(-)}]^2}
\frac{P_{\Omega -1}'(\xi)}{P_{\Omega}(\xi)}\right]
 +  4\rho_{12}^{2}\left[1-\xi\frac{P_{\Omega -1}(\xi)}
{P_{\Omega}(\xi)}\right]\right\}
\end{equation}
\narrowtext
are two contributions to the pairing energy (cf. Eqs.{{\ }}(A.7)
and (A.8) of Ref.{{\ }}\cite{NDS}), and
\begin{equation}
k=\rho_{22}-\rho_{11},
\end{equation}
\begin{equation}
\rho_{(\pm)}=\frac{1}{2}(1\pm\sqrt{k^{2}+4\rho_{12}^{2}}),
\end{equation}
\begin{equation}
\xi=\frac{\rho_{(+)}^{2}+\rho_{(-)}^{2}}{\rho_{(+)}-\rho_{(-)}}.
\end{equation}
In the above equations $P_{n}(x)$ is the Legendre polynomial of
the $n$-th order.  It is easy to see, that in the limit of very
weak pairing ($\xi\rightarrow$1, $P_n(1)$=1) the total pairing
energy given by Eqs.{{\ }}(\ref{epair1}) and (\ref{epair2})
becomes equal to $E_{\text{unpair}}$.  On the other hand, if
pairing is very strong [$\xi\rightarrow$$\infty$,
$P_{n-1}/P_n\rightarrow n/(2n-1)\xi^{-1}$,
$P'_{n-1}/P_n\rightarrow n(n-1)/(2n-1)\xi^{-2}$], the pairing
energy approaches the limit of the seniority model,
$-G\Omega(\Omega+1)$.

We have performed calculations within the
symmetric variant of the R(5) model for $e$=1,
$\Omega$=20 (i.e., the half--filled symmetric system
with
$n$=40 particles), and for three values of the pairing strength,
$G$=0.015, $G$=0.065, and $G$=0.1 \cite{units}. Without
rotation, the mean--field ({\sc bcs})
solution  undergoes a transition to the paired regime
at the critical strength $G$=2$e/(2\Omega-1)$$\simeq$0.051.
Therefore, the intermediate value of $G$=0.065 corresponds to
the phase transition region, and allows us to study the
destructive role of rotation on pairing correlations, while the
other two values of $G$ represent the weak and strong pairing
limits, respectively.

The results of the {\sc bcs} method are in
Figs.{{\ }}\ref{Fig3}--\ref{Fig4} denoted by the full
triangles
and full circles for
{\sc rbcs}
and  {\sc rfbcs}, respectively.
 The latter results are obtained by the
variation after projecting
the good particle number component
of the {\sc rbcs}
state.
   The results based on
the {\sc lnc} method are denoted by open
symbols.  The open triangles and open circles
represent the {\sc lnc}
and {\sc lnc}+{\sc pnp}
(exact particle number projection of {\sc lnc} solutions) results,
respectively.
The exact results are denoted by
the water--wheel symbols.

Calculations have been performed for
$\omega$ ranging from 0 to 1.2 \cite{units}.  For
each $\omega$, the energy and angular momentum have been
determined from Eqs.{{\ }}(\ref{etot}) and (\ref{itot}), or
Eqs.{{\ }}(\ref{etotpnp}) and (\ref{itotpnp}), and the pairing
energies have been computed
by means of Eq.{{\ }}(\ref{pair}).  In this way, the
plots of pairing energy versus angular momentum have been constructed and
are shown in Fig.{{\ }}\ref{Fig3}.

For the weak pairing strength, Fig.{{\ }}\ref{Fig3}a, the
{\sc rbcs} method gives only the unpaired solution for all spin
values \cite{BBD}.
 Although the solutions of the {\sc lnc} method contain some
pairing correlations, the approximate formula for the energy
with the corrective $\lambda_2$ term, Eq.{{\ }}(\ref{etot}),
gives the pairing energy much too small. On the other hand, when
the {\sc lnc} solutions are projected on the good particle
number one obtains a fair qualitative agreement with the exact
results (the maximum relative difference is of the order of
20\%).
This illustrates the fact that the
{\sc lnc} wave function describes resonable well the pairing
correlations in a weak pairing limit even if the average value
of the auxiliary Hamiltonian ${\hat K}^{\omega}$
is not a very good approximation to the exact energy of the system.

As seen in Fig.{{\ }}\ref{Fig3}a, the {\sc rfbcs} results
provide an excellent approximation to the
exact values.
This indicates,
that there is still some room for
improvements of the {\sc lnc}+{\sc pnp} method.
{}From
Ref.{{\ }}\cite{DN} it follows that the difficulties of the
{\sc ln} method in describing the half--filled ($n$=2$\Omega$)
system in the weak pairing limit stem from the fact that its
exact ground--state energy cannot be approximated by a
second--order expansion
in $n$ centered at $n_0$=2$\Omega$.  However,
the parabolic expansion works  very well for the
ground--state energies of {\em asymmetric}
systems with $n$$\ne$2$\Omega$.
Based on this result, a useful two--step procedure is suggested.
Firstly,
the {\sc ln} or {\sc lnc} equations are solved for the system with
$n_0$=$2\Omega\pm2$. In the second step,
the $n$=2$\Omega$ component is projected out from the
resulting {\sc lnc} wave
function.
Such a hybrid method relies
on {\em extrapolating} the $n$=2$\Omega$ solution from {\em
either} those for $n<$2$\Omega$ {\em or} those for $n>$2$\Omega$,
instead of {\em interpolating} between solutions for
$n<$2$\Omega$ {\em and} those for $n>$2$\Omega$.  Results of
{\sc ln}+{\sc pnp} calculations for $n_0$=2$\Omega$+2=42 are
presented in Fig.{{\ }}\ref{Fig6} and  agree remarkably well
with those
using the {\sc rfbcs} method and with the exact values.

The weak--pairing regime is realised in nuclei
around shell or subshell closures
where $e$$\gg$$G\Omega$. Therefore, the hybrid method described
above can be useful for studying properties of, e.g.,
spherical magic nuclei at low spin
or superdeformed
magic nuclei ($^{152}$Dy or $^{192}$Hg) at high spin.

For the intermediate pairing strength, Fig.{{\ }}\ref{Fig3}b,
the {\sc rbcs} method yields the unpaired solution above
a certain critical angular momentum. Such a sharp
transition is not present in the exact results.
 Although the low--spin
{\sc lnc} results are in much better
agreement with the exact solutions than those of the
{\sc rbcs} method, this method  yields too small pairing energy at
higher spins.  Not surprizing,
the pairing energies of
the {\sc pnp}+{\sc lnc} method
agree quite well with those obtained by means of the
{\sc rfbcs} treatment. The latter ones describes fairly well the exact
values, while the remaining small difference, which cannot be
accounted for by using the {\sc bcs}--type wave function,
illustrates the presence of large correlations
 (due to the quasiparticle interaction)
at the phase transition
region.

For the strong pairing limit, Fig.{{\ }}\ref{Fig3}c, the
{\sc rbcs} method is the only one which fails to describe exact
results. All other methods give very good agreement at low and
high angular momenta and leave a gap of missing solutions
around $I$=18, i.e., in the region where the
adiabatic cranking
approximation breaks down (the solutions in the phase transition region
correspond to the maximum of the total Routhian rather than to the
minimum \cite{NDS}).
Interestingly, the region of instability is very large for the {\sc rfbcs}
method, where it extends down to $I$=10, while the {\sc lnc}
method is able to follow solutions up to $I$=16. The particle
number projection does not lead to a significant improvement here.

Fig.{{\ }}\ref{Fig2} shows the pairing delta
$\Delta_{\text{BCS}}$ as a function of
$\omega$.  This quantity characterizes the {\sc bcs} states used
in {\sc rbcs}, {\sc lnc}, and {\sc rfbcs} methods and defines
the {\sc bcs} occupation probabilities
(for the {\sc ln} occupation probabilities, see, e.g.,  Ref. \cite{BHB}).
In fact, in the methods employing the concept of {\sc pnp},
$\Delta_{\text{BCS}}$ is not related to any particular observable;
it should not be understood as the energy gap, but rather as a
variational parameter (see Ref. \cite{NDS}).
Fig.{{\ }}\ref{Fig2}
shows
$\Delta_{\text{BCS}}$ together with
$\Delta_{\text{exact}}$$\equiv$${\sqrt{
-G\langle\hat{H}_{\text{pair}}\rangle}}$ where the average value
is calculated with respect to the exact ground-state wave
function. By definition,
$\Delta_{\text{exact}}$ is a direct measure of pairing correlations
and it should  reflect  $\Delta_{\text{BCS}}$ in the limit
of large pairing.
 Note that the $\Delta_{\text{BCS}}$ parameter for
{\sc lnc}+{\sc pnp} is the same as for {\sc lnc}, and, therefore,
is not shown in Fig.{{\ }}\ref{Fig2}.
It is seen in Fig.{{\ }}\ref{Fig2}a that
in the weak pairing limit the two quantities
$\Delta_{\text{exact}}$ and $\Delta_{\text{BCS}}$
differ very
much even if the {\sc rfbcs} pairing energy is rather close to
exact values, Fig.{{\ }}\ref{Fig3}.
In the intermediate and strong pairing limits the
$\Delta_{\text{BCS}}$ parameter of {\sc rbcs} goes to zero at
the critical angular momentum, which illustrates the
Mottelson--Valatin
phase transition.
The exact results do not show such a sharp transition, and the
$\Delta_{\text{BCS}}$ parameters of the {\sc lnc} and
{\sc rfbcs} method qualitatively reproduce the exact values.

The angular momentum alignment is illustrated in
Fig.{{\ }}\ref{Fig4} where the kinematical moment of inertia,
\begin{equation}\label{inert}
{\cal J}^{(1)}=\frac{I}{\omega_I},
\end{equation}
is shown as a function of rotational frequency $\omega_I$.
At this point
it
should be stressed
that the rotational frequency, $\omega_I$, obtained from
the canonical relation
\begin{equation}\label{omrot}
\omega_I = {{dE}\over{dI}}
\end{equation}
is equivalent to the cranking--model frequency $\omega$ only for
the exact eigenstates of a model cranking Hamiltonian.
Consequently, the relation
\begin{equation}\label{omrot1}
\omega=\omega_I
\end{equation}
holds  exactly
for the exact solutions of the R(5) model, and also for the
{\sc rbcs} variant (the solution is an eigenstate of the
independent--quasiparticle Hamiltonian).
On the other hand,
the relation (\ref{omrot1}) does not hold for the approaches
based on the {\sc pnp} treatment, since the resulting states
are determined from the {\em restricted} variational principle.
Consequently, the rotational frequency in
Fig.{{\ }}\ref{Fig4} is determined
using Eq.{{\ }}(\ref{omrot}).
The moment of
inertia,  Eq.{{\ }}(\ref{inert}), illustrates the
Mottelson--Valatin phase transition.
It can be seen that for all pairing strengths the
{\sc lnc} method correctly describes this transition, both with
and without the subsequent particle number projection.

We have also performed calculations for the
asymmetric system, $n$$\neq$2$\Omega$. Here, the
static {\sc rbcs}  solution is always present, independently
of the value of the pairing strength
\cite{BBD}, and the
results and conclusions are very similar to those for the
symmetric system, $n$=$2\Omega$, in the strong pairing limit.
The results for the $n$=48 system are shown in Fig.{{\ }}\ref{Fig7}.
It is seen that the {\sc lnc} method provides an excellent
approximation to the pairing energy, even without a subsequent
particle number projection.

\section{Conclusions}

We have studied the pairing correlations in rotating nuclei
using the cranked Lipkin-Nogami {\sc lnc} method which is based
on employing the auxiliary Hamiltonian
$\hat{K}^{\omega}=\hat{H}-\lambda_{1}\hat{N}-\lambda_{2}
\hat{N}^{2}-\omega\hat{j}_{x}$, where the parameters $\lambda_{1}$ and
$\lambda_{2}$ are chosen so that the influence of the nucleon
number fluctuation is strongly reduced. One should emphasize the
simplicity of the {\sc lnc} approximation, especially when
compared with more sophisticated projection methods.  In
practice, the {\sc lnc} method is a simple extension of the
usual {\sc rbcs} treatment.

Good accuracy was obtained for the ground state energy,
particularly in the case of strong pairing interaction.  It
means that the method suppresses correctly the ``dangerous"
(particle-number violating) part of the quasiparticle
interaction.  The weakness of the {\sc lnc} method in the weak
pairing limit can be overcome by performing the projection {\em
after} variation. Therefore,
the {\sc lnc+pnp} approach can provide
us with a fair description of rotating nuclei near
shell and subshell closures.

Another welcome feature of the {\sc lnc} method is that it
provides us with a very good description of the pairing phase
transition region, regardless of the strength of pairing
interaction.  Note that using the {\sc lnc} method one can obtain
non--trivial solutions even for very fast rotation where the
{\sc rbcs} method breaks down.

\acknowledgements
This research was supported in part by the Polish State
Committee for Scientific Research under Contract
No.{{\ }}20450~91~01.  The Joint Institute for Heavy Ion
Research has as member institutions the University of Tennessee,
Vanderbilt University, and the Oak Ridge National Laboratory; it
is supported by the members and by the Department of Energy
through Contract Number DE--FG05--87ER40361 with the University
of Tennessee.  Theoretical nuclear physics research at the
University of Tennessee is supported by the U.S. Department of
Energy through Contract Number DE--FG05--93ER40770.

\begin{figure}\caption{
Energy levels for the R(5) model.  The above pattern is repeated
$\Omega$ times.
\label{Fig1}}
\end{figure}

\begin{figure}\caption{
Pairing energy versus angular momentum for three values of
pairing strength, $G$=0.015 (a), $G$=0.065 (b), and $G$=0.1 (c).
\label{Fig3}}
\end{figure}

\begin{figure}\caption{
Pairing energy versus angular momentum for the weak pairing
strength, $G$=0.015.  The {\sc lnc}+{\sc pnp} results calculated
for $n$=$2\Omega$=40 from the $n_0$=42 solutions are also shown.
\label{Fig6}}
\end{figure}

\begin{figure}\caption{
Pairing delta  $\Delta$ versus rotational frequency $\omega_I$ for
three values of  pairing strength, $G$=0.015 (a), $G$=0.065 (b),
and $G$=0.1 (c).
\label{Fig2}}
\end{figure}

\begin{figure}\caption{
Kinematical moment of inertia versus rotational frequency
$\omega_I$ for three values of  pairing strength, $G$=0.015 (a),
$G$=0.065 (b), and $G$=0.1 (c).
\label{Fig4}}
\end{figure}

\begin{figure}\caption{
Pairing energy versus angular momentum for the asymmetric
variant of the R(5) model with $n$=48 and
for three values of  pairing strength, $G$=0.015 (a),
$G$=0.065 (b), and $G$=0.1 (c).
Only the exact,
{\sc rbcs} and {\sc lnc} results are displayed.
\label{Fig7}}
\end{figure}


\begin{references}
\bibitem {units} All energies are supposed to be expressed in
                 the same units, say MeV's, and the units are
                 not shown explicitly.
\end{references}
\end{document}